\documentclass[sigconf,]{acmart}

\usepackage{booktabs}                  
\usepackage{hyperref}
\usepackage{enumitem}
\setlist{topsep=0pt, leftmargin=*}

\definecolor{revision}{RGB}{255,127,14}
\definecolor{forestgreen}{RGB}{50,120,50}
\newcommand{\revision}[1]{\textcolor{revision}{#1}}

\newcommand{\etal}{et al.}
\newcommand{\eg}{e.g.,\ }

\newcommand{\bstart}[1]{\vspace{1mm} \noindent{\textbf{#1.}}}

\newcommand{\bdef}[1]{\vspace{1mm} \noindent{\textbf{#1}}}

\newcommand{\brian}[1]{\textcolor{forestgreen}{(BH: #1)}}
\newcommand{\lyn}[1]{\textcolor{purple}{(LB: #1)}}
\newcommand{\matt}[1]{\textcolor{blue}{(MB: #1)}}
\renewcommand{\revision}[1]{#1}
\renewcommand{\brian}[1]{}
\renewcommand{\lyn}[1]{}
\renewcommand{\matt}[1]{}

\AtBeginDocument{%
  \providecommand\BibTeX{{%
    \normalfont B\kern-0.5em{\scshape i\kern-0.25em b}\kern-0.8em\TeX}}}

\copyrightyear{2022}
\acmYear{2022}
\setcopyright{iw3c2w3}
\acmConference[UIST '22]{The 35th Annual ACM Symposium on User Interface Software and Technology}{October 29-November 2, 2022}{Bend, OR, USA}
\acmBooktitle{The 35th Annual ACM Symposium on User Interface Software and Technology (UIST '22), October 29-November 2, 2022, Bend, OR, USA}
\acmDOI{10.1145/3526113.3545614}
\acmISBN{978-1-4503-9320-1/22/10}

\acmSubmissionID{6449}

\begin{document}


\title{Augmented Chironomia for Presenting Data to Remote Audiences}

\author{Brian D. Hall}
\affiliation{%
  \institution{University of Michigan}
  \city{Ann Arbor}
  \state{Michigan}
  \country{USA}
}
\email{briandh@umich.edu}

\author{Lyn Bartram}
\affiliation{%
  \institution{Simon Fraser University}
  \city{Surrey}
  \state{British Columbia}
  \country{Canada}}
\email{lyn@sfu.ca}

\author{Matthew Brehmer}
\affiliation{%
  \institution{Tableau Research}
  \city{Seattle}
  \state{Washington}
  \country{USA}}
\email{mbrehmer@tableau.com}

\renewcommand{\shortauthors}{Hall et al.}

\begin{abstract}
To facilitate engaging and nuanced conversations around data, we contribute a touchless approach to interacting directly with visualization in remote presentations. We combine dynamic charts overlaid on a presenter's webcam feed with continuous bimanual hand tracking, demonstrating interactions that highlight and manipulate chart elements appearing in the foreground. These interactions are simultaneously functional and deictic, and some allow for the addition of ``rhetorical flourish'', or expressive movement used when speaking about quantities, categories, and time intervals. We evaluated our approach in two studies with professionals who routinely deliver and attend presentations about data. The first study considered the presenter perspective, where 12 participants delivered presentations to a remote audience using a presentation environment incorporating our approach. The second study considered the audience experience of 17 participants who attended presentations supported by our environment. Finally, we reflect on observations from these studies and discuss related implications for engaging remote audiences in conversations about data.
\\
\\
\\
\end{abstract}



    \begin{teaserfigure}
        \centering  
        \vspace{-2.5mm}
        \includegraphics[width=\linewidth]{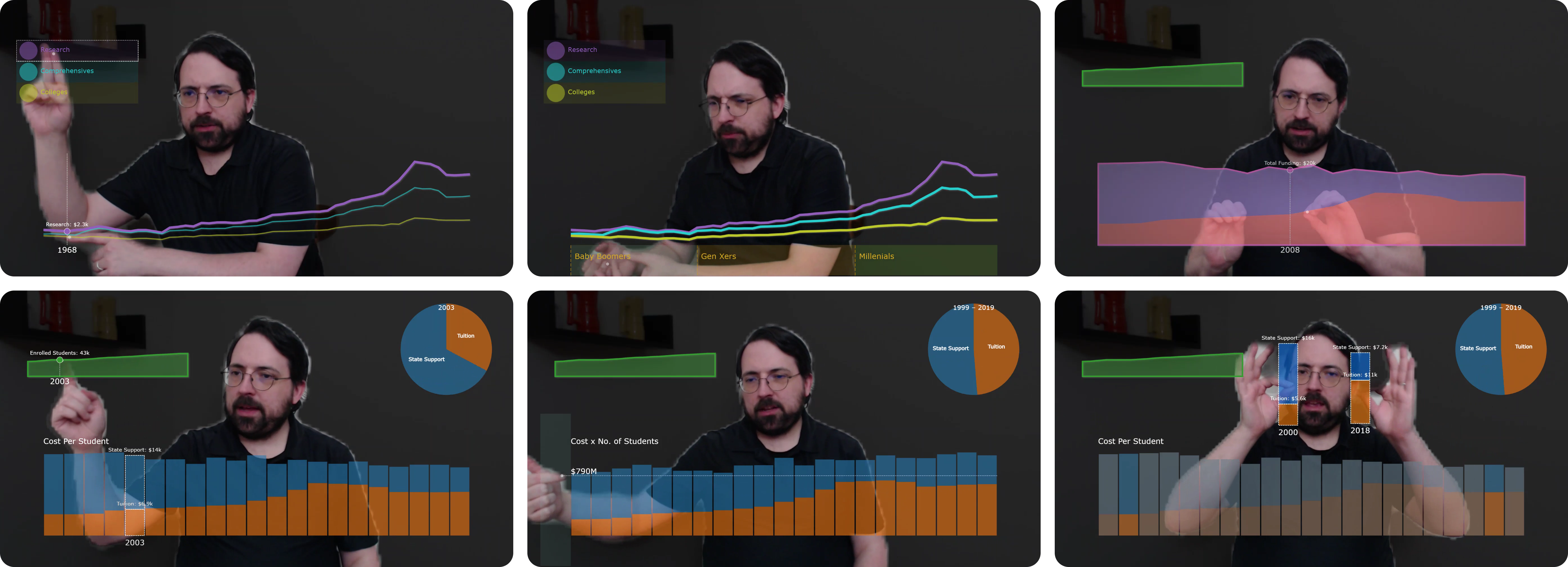}
        \vspace{-2.5mm}
        \caption[Frames from a video presentation in which a presenter manipulates and highlights elements in chart overlays.]{Frames from a video presentation in which a presenter manipulates and highlights elements in chart overlays.}
        \label{fig:teaser}
    \end{teaserfigure}

\maketitle

\section{Introduction}
\label{sec:intro}

\textit{Chironomia} is the ancient art of manual rhetoric~\cite{austin1806chironomia}: it refers to expressive hand movements that a speaker can employ when communicating to an audience.
In this paper, we consider the combination of this art with the rhetorical power of data visualization~\cite{hullman2011visualization}.
We demonstrate an augmentation of webcam video with interactive visualization overlays in a speaker's foreground (Figure \ref{fig:teaser}).
We further augment a speaker's hand movement, granting them functional control of the overlays without precluding the potential for illustrative and affective body language. 
Together, these augmentations allow for nuanced and engaging conversations about data with remote audiences.

Our work is primarily motivated by the ever-growing needs of people within organizations who lead discussions and inform decisions that are grounded in data~\cite{brehmer2021jam,dimara2021unmet}.
These activities manifest in synchronous meetings and presentations among and between colleagues, stakeholders, executives, and customers.
In these settings, participants support their discussions and arguments with data, using charts and graphs to express quantities, proportions, categories, and time intervals.
These visual aids are typically presented as a series of static slides featuring a small palette of familiar chart types~\cite{brehmer2021jam}.
\revision{Typically, people paste chart screenshots on to slides and add annotations or shape masks that draw attention to specific values or visual patterns; these attentional cues can only be revealed or removed in a pre-planned sequence during presentation delivery. 
If the underlying screenshots need to be updated or replaced, the attentional cues must also be manually updated to match the new content. 
Alternatively, a presenter could prepare multiple screenshots of a chart in various states, so as to reveal, emphasize, and hide specific elements, but preparing, organizing, and staging these assets is tedious.}
Part of the problem we address in this work is the inability to interact with such content in unplanned ways, such as in response to audience questions.

The other part of the problem is the increased prevalence of remote and hybrid work environments since the onset of the COVID-19 pandemic.
Given this shift, presentations are now predominantly supported by the screen-sharing functionality of teleconference applications. 
This impoverished presentation medium reduces the capacity for presenters to draw their audience's attention to specific data points and fully illustrate their interpretations of the data.
Absent the affective and illustrative cues that a presenter can provide when co-located with an audience, they cease to become `sage on the stage'; instead, they are relegated to a secondary thumbnail video frame or just a disembodied voice.
As a result, audiences become disengaged and opportunities for discussion are lost.


The approach we take in this paper was inspired in part by compelling public-facing presentations of data that take advantage of multiple communication modalities: the oration and body language of an engaging presenter coupled with dynamic data visualization.
In particular, we draw inspiration from presentations delivered by the late public health expert Hans Rosling~(\eg~\cite{Rosling2006,Rosling2007,Rosling2010,Rosling2013}). 
In his 2010 BBC Documentary \textit{200 Countries, 200 Years, 4 Minutes}~\cite{Rosling2010}, an animated scatterplot was composited over a recorded video of him speaking and gesticulating, giving the impression that he was controlling the chart with his body.
In this paper, we demonstrate that it is now possible to deliver such presentations to a synchronous audience by compositing interactive charts over conventional webcam video.
This approach allows a more attentive audience to interrupt the presenter with questions, prompting unplanned interactions with the data.



\bstart{Contributions}
We contribute a functional realization of a presentation environment in which live webcam video is overlaid with interactive visualization controlled by a presenter's hand movements. 
We also contribute demonstrations of several categories of touchless  interaction that are simultaneously deictic and functional, without foreclosing the potential for affective gesture and rhetorical flourish.
Finally, we contribute reflections and observations from two formative studies of our approach with 29 professionals who routinely attend and deliver presentations about data, considering both presenter and audience perspectives.
\section{Background and Related Work}
\label{sec:rw}

We build upon several bodies of knowledge:
classifications of hand movement in multimodal communication, gestural and spatial interaction, and the communicative use of data visualization. 

\subsection{Talking with the Hands}
\label{sec:rw:gesture}

The non-verbal communication modalities of bodily movement, posture, gaze, and facial expression are interwoven with speech in social interaction
~\cite{studdert1994hand,verhulsdonck_virtual_2009}. 
The communicative use of body language has been documented well into antiquity~\cite{austin1806chironomia}, 
with hand gesture in particular being viewed as integral to interpersonal communication~\cite{maurer_exploring_2017}.  
According to a co-speech perspective~\cite{wagner_gesture_2014,dohen_co-production_2017}, gestures amplify and extend the production and perception of meaning, and 
they contribute to discourse coherence when speakers supplement and extend the interpretation of prior gestures and speech~\cite{lascarides_discourse_2009}.

An embodied cognition perspective~\cite{matsumoto_nonverbal_2013} posits that hand gestures are critical for comprehending and reasoning about complex content, such as
when teaching mathematical concepts~\cite{edwards_gesture_2005,bjuland_interplay_2008,aldugom_gesture_2020} or when establishing a shared understanding in engineering and design~\cite{cash_prototyping_2016,streeck_gesturecraft_2009}.
Complex content also arises in persuasive presentations given by entrepreneurs and investors, and presenters who use their hands to depict and symbolize business ideas tend to have more positive entrepreneurial outcomes relative to those who present via the modalities of speech and text alone~\cite{cornelissen_sensegiving_2012,clarke_actions_2019}. 
There is consistency in hand gesture when speaking about quantities in particular; a study of television news archives~\cite{winter2013using} showed that people tend to compress or expand their fingers or hands when speaking about small and large values, respectively, or to move their hands laterally, vertically, or distally to indicate a progression of values.
Given that many types of charts shown in presentations depict not only quantities but also time intervals and categories, our work seeks to better understand the role of expressive hand movements performed in the presence of these visual aids.


In considering prior categorizations of communicative hand movement~\cite{matsumoto_nonverbal_2013,studdert1994hand}, we see two recurring categories as being particularly relevant to scenarios where people speak about data:

\bdef{Illustrative} gestures supplement what is being said, with
three subcategories of interest: \textit{\textbf{deictic}} gestures draw attention to artifacts visible to both speaker and audience through pointing, enumerating, and framing; \textit{\textbf{iconic}} gestures illustrate distances, sizes, and shapes; and \textit{\textbf{metaphoric}} gestures communicate abstractions~\cite{clarke_actions_2019}, such as moving the hand clockwise to signify the passage of time.

\bdef{Affective} gestures convey emotion and emphasis; these \textit{performative} gestures include the tightening of a fist or throwing one's hands up in the air to convey uncertainty or dismay. 
Consider how a \textit{beat} of the finger, hand, or arm can mark an important point in a speech, while a repeated beat can variably convey urgency, importance, or steadiness, depending on the tempo.
From the ancient Greeks to modern public speaking coaches, many have considered the richness of this category; 
Cicero used the term \textit{chironomia}~\cite{austin1806chironomia} to describe gesticulation in the service of rhetorical delivery, argumentation, and persuasion~\cite{smith2019}.
It is therefore no surprise that political speeches employing affective hand gestures are perceived as being more compelling, persuasive, and engaging~\cite{bull_use_1986}. 

Given the persuasive~\cite{pandey2014persuasive} and rhetorical~\cite{hullman2011visualization,kosara2017argument} potential of data visualization, our research considers scenarios where an audience can simultaneously see visual representations of data with an orator performing illustrative and affective hand gestures, movements that may also take on a \textit{functional} role, which we discuss next.

\subsection{Gestural Interfaces}
\label{sec:rw:presentation}

HCI research spanning multiple decades has focused extensively on the technical and \textit{functional} aspects of gestural interaction, such as how to control an interface using gesture (\eg see Bolt (1980)~\cite{bolt1980}). 
While touch-based gestures common to mobile computing are now ubiquitous~\cite{cirelli_survey_2014} (\eg pinch, swipe, tap and hold), the lexicon of touchless or mid-air interaction is still evolving~\revision{\cite{aigner2012understanding}}, \revision{particularly since the introduction of commercially-available visual sensing interfaces such as the Microsoft Kinect and the Leap Motion Controller.}
Despite the growing body of research dedicated to this interaction modality, \revision{it largely reflects a focus on recognition and usability~\cite{ohara_interactional_2014,vuletic_systematic_2019}}.
This research highlights issues of how common gestural techniques can both enhance and interfere with the contexts in which they are used. 
\revision{For example, O'Hara~\etal~\cite{ohara_interactional_2014} demonstrated a touchless interface for surgeons to control the display of medical imagery while performing surgery; while surgery is a collaborative context, the functional gestures considered were intended primarily for system control rather than for communicating with other members of the surgical team.  
This example underscores the need for a} unified framework that encompasses both the \textit{functional} and \textit{communicative} power of touchless interaction with the hands.

\bstart{Gesture-controlled presentations}
According to Harrison~\cite{harrison_showing_2021}, speakers giving presentations supported by visual aids such as slides produce complex multimodal `ensembles' comprised of speech, body language, and images, where one element in an ensemble may emphasize, reinforce, or restate the meanings of other elements. 
When presenting visuals to a co-located audience, Fourney~\etal~\cite{fourney2010gesturing} found that people 
employ various deictic gestures with either hand to variably emphasize different content groupings, from `everything' to specific visual elements. 

We are aware of several precedents for presentation systems incorporating touchless interaction with the hands~ \revision{\cite{baudel1993charade,lee_hmm-based_1999,kim2018holobox,fourney2010gesturing,matulic2016embodied,cuccurullo_gestural_2012,saquib_interactive_2019}}, including sensor-based~\cite{baudel1993charade} and vision-based systems~\revision{\cite{fourney2010gesturing,lee_hmm-based_1999,kim2018holobox,harika_finger-pointing_2016,matulic2016embodied}}. 
Collectively, these projects document appropriate algorithms to effectively capture a set of gestures that control a progression of visual aids~\cite{baudel1993charade,lee_hmm-based_1999,harika_finger-pointing_2016} as well as considerations of interaction learnability and memorability~\cite{cuccurullo_gestural_2012}. 
While those who used these systems generally found their interactive experiences to be novel and engaging, the scope of interaction was limiting~\cite{fourney2010gesturing,baudel1993charade, cuccurullo_gestural_2012}, in that system designers prioritized slide navigation over interacting with slide content~\cite{fourney2010gesturing}.
Additional issues included the capture of unintentional movements~\cite{fourney2010gesturing, cuccurullo_gestural_2012} and an increased cognitive load when the repertoire of interactions was large and unconstrained~\cite{saquib_interactive_2019}. 

We specifically call attention to the recent work by Saquib~\etal~\cite{saquib_interactive_2019}, who demonstrated a Kinect-based system for authoring creative video storytelling performances.
Using their system, a performer could manage video navigation as well as point at and modify the appearance of animated graphical elements in the foreground via mid-air gesture. 
However, the interactions and associated behavior of linked visual elements had to be defined when authoring; while those who used the system explored a variety of interactions when authoring, they executed fewer interactions at performance time, and these interactions had to be performed in a predictable way.

In contrast to previous systems, our approach incorporates a standard webcam for tracking hand movements, and we focus on the small yet ubiquitous palette of chart types that are common across presentations of data~\cite{brehmer2021jam}.
Moreover, we demonstrate categories of interaction that can be performed flexibly across chart types during a presentation, and we prioritize the presentation delivery and audience experiences over the authoring experience.





\bstart{Gestural interaction with data}
Prior research has also considered the potential of touchless interaction for exploratory data analysis.
For instance, in virtual reality environments, systems like ImAxes~\cite{cordeil2017imaxes} allow people to move and coordinate three-dimensional chart components within a virtual environment.
Elsewhere, we have seen mobile augmented reality applications allow for interaction with virtual objects with the dominant hand while the non-dominant hand holds the mobile device~\cite{issartel2014slicing,qian2019}.
Altogether, this body of work elicits the question of what happens when an audience is watching someone interact with data in these ways and the extent to which hand movement is interpreted as being communicative.
\subsection{The Communicative Use of Visualization}
\label{sec:rw:storytelling}

Apart from data analysis, people also visualize data to communicate with others: it can help them tell stories about data~\cite{riche2018data}, support an argument~\cite{kosara2017argument}, or persuade an audience~\cite{pandey2014persuasive} to make decisions that are grounded in data~\cite{dimara2021unmet}.
Narrative visualization~\cite{segel2010narrative} can assume a variety of forms and rhetorical structures~\cite{hullman2011visualization}, though much of the prior research in this area has focused on forms typically associated with web-based journalism, where dynamic visualization manifests in magazine-style articles \cite{conlen2018idyll,mckenna2017visual}, reader-controlled slide presentations \cite{boy2015storytelling,satyanarayan2014authoring}, or recorded data videos~\cite{amini2015understanding}.
Each of these forms entail asynchronous consumption by individual viewers, and while they may be able to interact with the content, there is little capacity for them to interact with content authors or other viewers~\cite{mcinnis2020rare}.
In contrast, we focus on synchronous multimodal communication involving visualization delivered by a presenter.

Meanwhile, the prevalence of visualization in live television broadcasts has increased in recent years, a medium that has been largely ignored by the research community~\cite{drucker2018}. 
From weather reporting to coverage of the COVID-19 pandemic or the 2020 US Federal Election, correspondents have made use of large high-resolution touchscreen displays or those controlled by a handheld tablet.
We have also seen more elaborate presentations that place correspondents in an augmented reality environment produced with specialized cameras and studios (\eg~Vizrt~\cite{vizrt2021}). 
While these are live presentations about data, we focus on scenarios where a direct engagement with the audience is possible, and we assume no specialized equipment aside from a standard webcam.

\bstart{Conversations about data with a live, yet remote audience}
We concentrate on teleconference-based presentation settings where the speaker and audience are able to interact in real-time.
While prior research has demonstrated purpose-built tools for using dynamic data visualization in presentations for a co-located audience (such as SketchStory~\cite{lee2013sketchstory} or SandDance~\cite{drucker2015unifying}), 
there is a dearth of analogous purpose-built tools for live presentations about data to remote audiences~\cite{zhao2022stories}.
This absence is sorely felt in informal presentations taking place within (distributed) organizations~\cite{brehmer2021jam}, where interruptions and impromptu discussion about the specifics of the data are commonplace. 
Currently, presenters resort to screen sharing slides or visual analysis tools, such as analytical notebooks or business intelligence dashboards.
Slides impart an inappropriate level of formality and linearity, precluding spontaneous shifts of audience attention, while visual analysis tools introduce visually distracting interfaces, reflecting their intended use in individual interactive analysis.
Our work is a response to this gap in the tool landscape for presenting data.

\section{Process and Implementation}
\label{sec:proc}
Our goal was to develop a better experience for presenters and remote audiences alike, encouraging a level of audience engagement that is lost in remote communication, one that fosters nuanced conversations about data.   
This goal shaped our first design imperative: to support a presenter's direct interaction with the \textit{content} of the presentation and to look beyond purely functional view navigation and passive highlighting.
Our second design imperative was to demonstrate a presentation environment that would incur little to no cost or special equipment while enabling presenters' existing communication skills, namely their fluency with respect to performing illustrative and affective gestures.



\bstart{Iterative design}
We iteratively developed our approach to presenting data.
Our \revision{first proof-of-concept~\cite{infohands2021}} made use of greenscreen-compatible presentation tools (OBS~\cite{obs2022} and mmhmm~\cite{mmhmm2022}) along with a pose recognition model that we created using Teachable Machine~\cite{teachablemachine} to trigger transformations to a chart composited behind a presenter.
However, from an audience perspective, the poses seemed stilted, and echoing findings from Fourney~\etal~\cite{fourney2010gesturing}, it was undesirable for the presenter to occlude the content.
From a presenter perspective, a choreographed performance of poses was fatiguing.
Moreover, a set of discrete poses was insufficient for drawing attention to data; for instance, we could not use them to select or highlight specific chart elements.
Our second proof-of-concept was inspired by a news anchor sitting at a desk, with `over-the-shoulder' graphics composited on top of their video.
However, rather than composite charts with opaque backgrounds over webcam video, we lowered the opacity of chart elements and removed their backgrounds, so as to allow a presenter to point to chart elements from behind, reminiscent of the Lucid Touch interface~\cite{wigdor2007}.
In addition to semi-transparent over-the-shoulder charts, we incorporated continuous hand and finger tracking through the use of MediaPipe~\cite{mediapipe2020}.
However, like our earlier greenscreen prototype, interacting repeatedly with over-the-shoulder charts was physically awkward and fatiguing.
\bstart{An integrated browser-based presentation environment}
We retained the continuous hand tracking from our second prototype, integrating it into a browser-based presentation environment developed as a Svelte~\cite{svelte2022} application. 
\revision{We rendered the overlay charts rendered as SVG elements, and we made use of D3.js~\cite{bostock2011d3} scale transformations for both element placement and interaction handling.}

Following an observation that most presentations about data taking place within organizations incorporate a handful of simple chart types~\cite{brehmer2021jam}, our presentation environment is \revision{currently supports} variants of bar, line, area, and pie charts.
\revision{However, our environment can be easily extended to accommodate any SVG-based chart.}

Realizing the limitations of over-the-shoulder charts and taking inspiration from Rosling's 2010 documentary~\cite{Rosling2010}, our environment allows for chart overlays to fill the entire video frame.
This flexibility gives presenters the freedom to sit stationary or to stand and walk around the frame; it also allows them to keep their hands in more comfortable positions when speaking.
\revision{Additionally}, as presenters' surroundings and lighting arrangements will vary, we optionally apply background segmentation to darken the presenter's surroundings (Figure \ref{fig:teaser}) as well as a grayscale filter to the presenter's video, so as to place additional emphasis on the overlays.

\revision{Given the stateless nature of the MediaPipe API~\cite{mediapipe2020}, our implementation adds a time-based event protocol for gesture detection with dwell and timeout durations ranging between 0.1 to 1.0s; these settings reduce jitter as well as false positive and false negative classifications. 
For graphical performance, we avoid overprocessing the video stream, thereby reducing presenter-side latency to imperceptible levels. 
Specifically, we only send an image frame to MediaPipe for landmark detection if a previous frame is not being processed, and we perform gesture detection only once per result received from MediaPipe.}

Prior to giving a presentation with this environment, a presenter needs to specify \textit{scenes}, where each scene can contain one or more chart overlays connected to local data sources.
For each overlay, the presenter can enable its ability to respond to interaction and specify its visibility, dimensions, and positions within the scene.

We reiterate that our focus in this paper is not the presentation authoring experience; while the current scene and overlay specification is JSON-based, future authoring experiences could emulate drag-and-drop dashboard creation in business intelligence tools.

\bstart{\textit{What you see is what I see}}
Initially, we had intended for presenters to perform both communicative hand movements as well as those dedicated to scene management, such as adding, removing, positioning, and resizing overlays.
However, scene management would require either a set of unique functional gestures or an on-screen mode-switching widget. 
Both approaches would be distracting for audiences, particularly as neither serve a direct communicative purpose.
Furthermore, as with interacting with over-the shoulder overlays, scene management gestures can be tiring, echoing reports of actors becoming easily fatigued while filming similar touchless interactions in science fiction films~\cite{Noessel2012}.
Alternatively, we considered a secondary presenter view, such as a control panel not visible to the audience, but this would run the risk of splitting the presenter's attention, requiring them to coordinate interaction across two displays.
As a result, the presenter and audience both see the same content, and we relegated scene navigation in our environment to off-screen keyboard shortcuts that trigger animated transitions.
Chart overlays can enter or exit through translation or fade transitions, or if two consecutive scenes contain the same overlay, it can smoothly translate and scale if required.
\revision{Combining a \textit{`what you see is what I see'} approach~\cite{stefik1987wysiwis}} with keyboard-based scene navigation allowed us to focus on a presenter's communicative movement, which we discuss in the next section.

    \begin{figure*}[h!]
        \includegraphics[width=\linewidth]{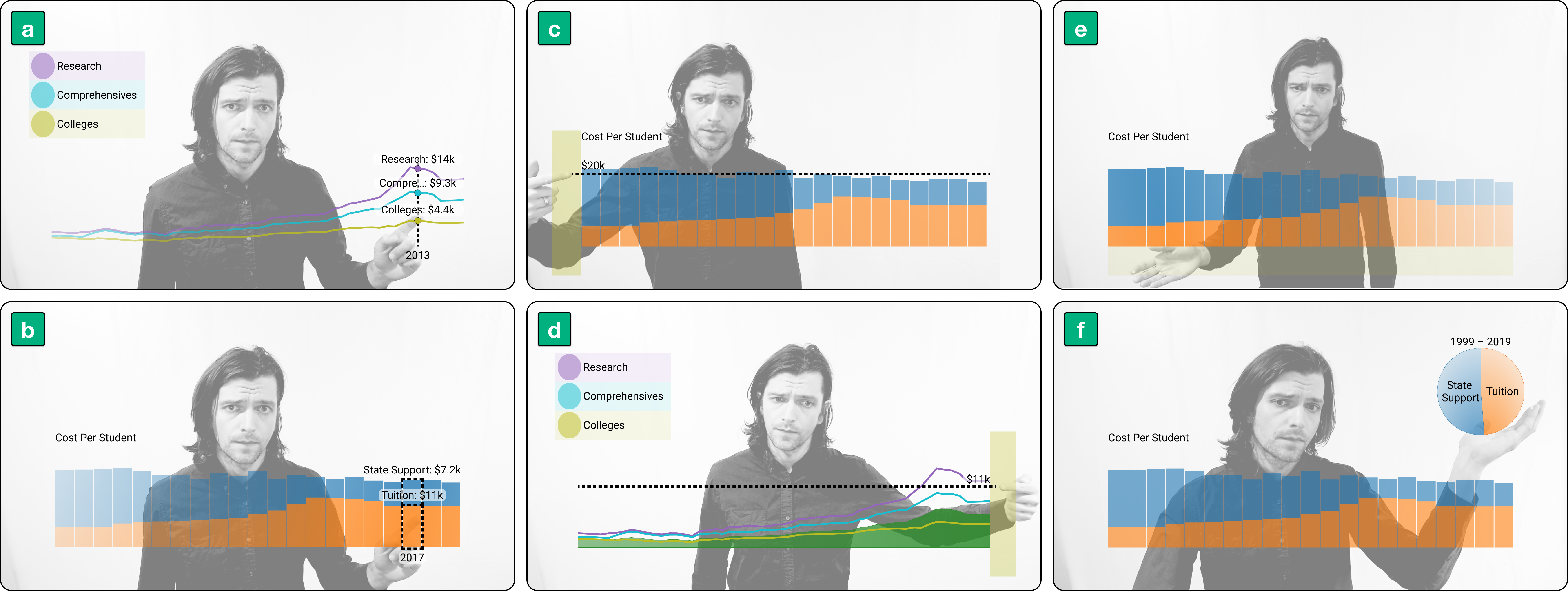}
        \vspace{-5mm}
        \caption{Three interactions applicable across chart types: (a---b) pointing with the index finger highlights the nearest value(s); (c---d) pointing with the index finger in either side margin triggers a reference line and value annotation; (e---f) an opacity gradient radiates from the palm's centroid to coarsely emphasize content near the hand.}
        \label{fig:palmpointmargin}
        \vspace{-2.5mm}
    \end{figure*}

    \begin{figure*}[h!]
        \includegraphics[width=\linewidth]{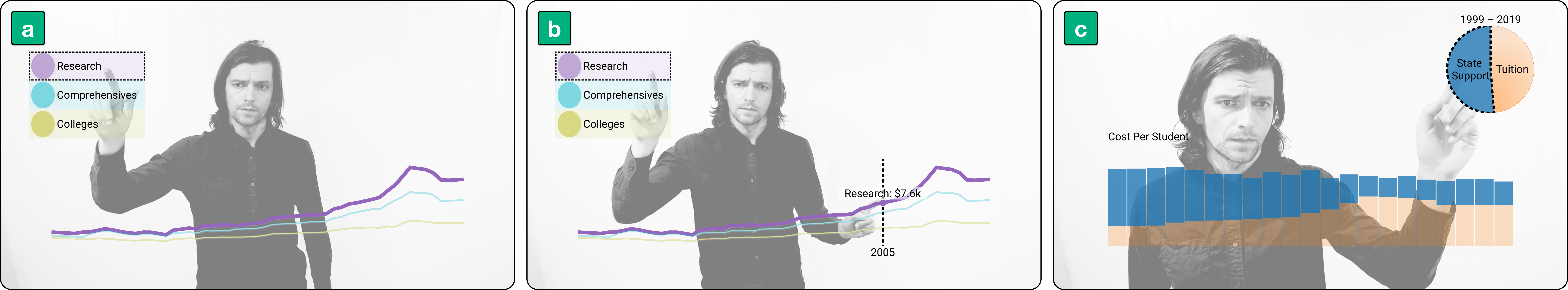}
        \vspace{-5mm}
        \caption{Highlighting across overlays: (a) pointing to a legend swatch emphasizes the corresponding series in an adjacent line chart; (b) if pointing at a legend swatch, only values corresponding to that series will appear as annotations (compare to Figure \ref{fig:palmpointmargin}a); (c) pointing to a wedge in a pie chart emphasizes marks having that category in an adjacent bar chart.}
        \label{fig:legend}
        \vspace{-2.5mm}
    \end{figure*}

    \begin{figure*}[h!]
        \includegraphics[width=\linewidth]{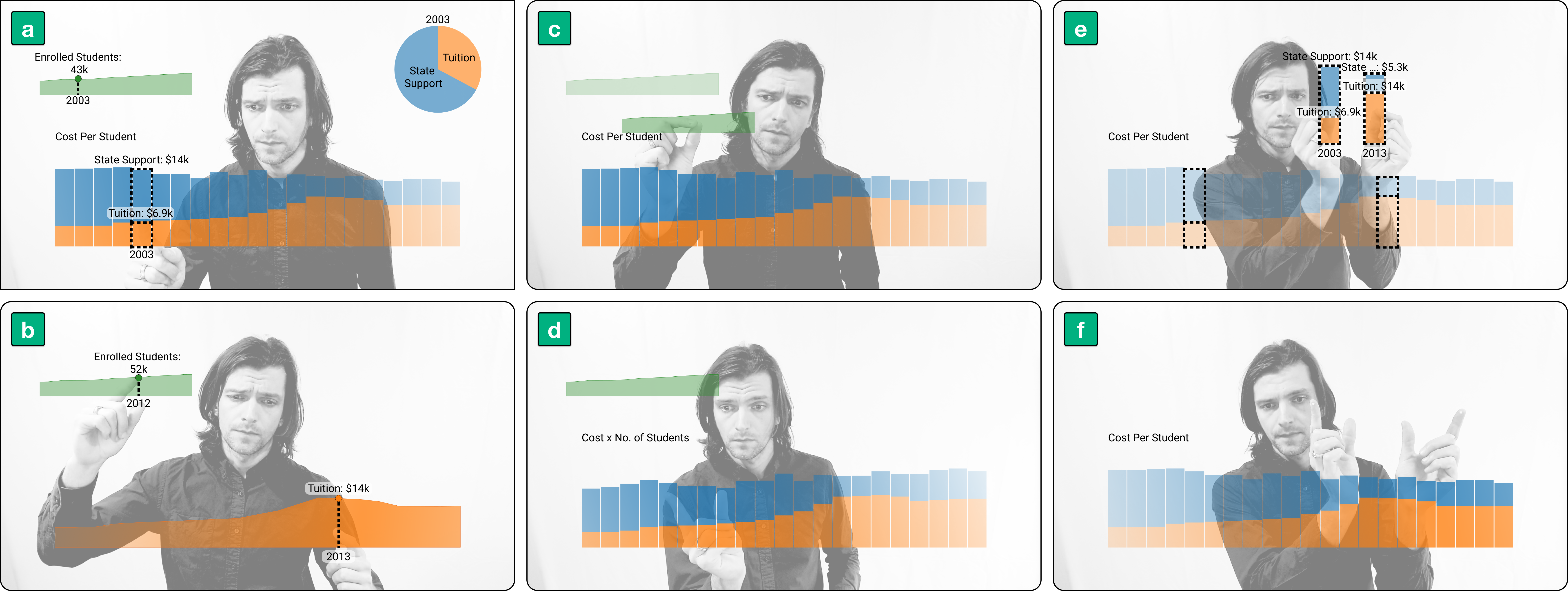}
        \vspace{-5mm}
        \caption{Coordinating and manipulating overlays: linked highlighting across charts (a) can be superseded with (b) bimanual pointing; (c) pinching and dragging a cloned copy of the green student enrollment area chart; (d) releasing the cloned area chart over the cost-per-student bar chart (which shares the same temporal domain) multiplies corresponding values, yielding total cost values over time; (e---f) pinching generates ephemeral and draggable cloned elements for spatially-adjacent comparisons.}
        \label{fig:multichart}
        \vspace{-2.5mm}
    \end{figure*}


\section{Interacting with Chart Overlays}
\label{sec:ixd}

We now describe the categories of chart overlay interactions supported by our environment.
In addition to their depictions in Figures~\ref{fig:palmpointmargin} through~\ref{fig:snapcrimp}, the supplemental video includes demonstrations of interactions from each category.
Overall, these touchless interactions are simultaneously deictic and functional, designed with the goal of avoiding conflict with affective movement. 
Most are agnostic to chart type, while some are unique to the type of data being shown or the geometry of a particular visual encoding. 
While interactions can be performed with a single chart overlay, we also show their effects with multiple adjacent or layered overlays. 
Many allow for the addition of \textit{rhetorical flourish}, or expressive movement that reinforces what is being said to advance a narrative about data.




Our environment tracks both of the presenters' hands, specifically the tip of the index finger, the tip of the thumb, and the palm centroid of both hands. 
\revision{The environment can optionally highlight index finger and thumb control points, and these markers will change color when the presenter is pinching.}
It can also distinguish between right and left hands, and the two hands can perform different interactions concurrently.
When both hands are within the same chart overlay, the right hand is designated as dominant and given precedence for single-handed interactions; however, it is possible to configure the environment for left-handed individuals.

\subsection{Deictic Highlighting}
\label{sec:ixd:deictic}

Figure \ref{fig:palmpointmargin} demonstrates three categories of deictic highlighting applicable across chart types, drawing inspiration from previous tools for emphasizing chart elements such as SmartCues~\cite{subramonyam2018smartcues} and ChartAccent~\cite{ren2017chartaccent}.
These interactions are continuous and ephemeral, updating based on the position of the fingertip or palm, with highlights disappearing once the hand leaves the bounds of an overlay.

\bstart{Pointing to reveal values}
Figure \ref{fig:palmpointmargin}a---b illustrates overt highlighting via pointing.
This form of highlighting directs attention to specific values in a chart, so we paired this interaction with tooltip-like value annotations.
For rectilinear charts with continuous scales along the horizontal axis (such as line and bar charts), we map the position of the index finger to the nearest bisected value in the underlying data; we display those values as text labels along with a vertical reference line and an emboldened mark stroke.
For circular charts (such as a pie chart), we use a trigonometric function to determine the segment nearest the index finger (Figure \ref{fig:legend}c).

\bstart{Pointing in the margin}
Figure \ref{fig:palmpointmargin}c---d shows the highlighting of an axis value spanning the entire chart. 
Whenever an index fingertip is found within the side margins of a rectilinear chart, we draw a horizontal reference line emanating from the fingertip.
This form of highlighting is both deictic and iconic, in that the raising of a hand tends to signify a larger quantity.

\bstart{Illuminating from the palm}
Figure \ref{fig:palmpointmargin}e---f shows a more subtle form of highlighting, where a radial opacity gradient emanates from the centroid of the palm.
This form of highlighting is intended to draw the audience's attention to a general region within a chart. 
This gradient can be activated whenever the presenter's palm appears within a chart or its margins; it is not activated when the palm is kept outside the margin (\ref{fig:palmpointmargin}c---d).
A presenter can therefore selectively combine this coarse highlighting with finer highlighting by intentionally placing the palm relative to the index finger.
If the palms of both hands are visible within a single chart, the gradient will emanate from the dominant hand.

\subsection{Interacting with Multiple Overlays}
\label{sec:ixd:multiple}

The display of multiple adjacent or superimposed overlays offers an opportunity to coordinate interaction between them.
While there are many chart coordination design patterns to consider~\cite{chen2020composition}, we predominantly focus on linked highlighting and selection.

\bstart{Highlighting categories}
Figure \ref{fig:legend} illustrates linked highlighting triggered by pointing at a categorical legend overlay (\ref{fig:legend}a), whereupon the corresponding series in the adjacent line chart is emphasized and other series are de-emphasized. 
At this point, the right hand can point at features in the line chart (\ref{fig:legend}b) while continuing to point at a particular legend swatch with the left hand; doing so will suppress the value annotations of other categories (compare to Figure \ref{fig:palmpointmargin}a).
Finally, linked highlighting need not be driven from a dedicated categorical legend: 
Figure \ref{fig:legend}c illustrates how pointing at a category in a pie chart emphasizes the corresponding category in an adjacent bar chart. 

\bstart{Highlighting values along a common domain}
Linked highlighting can also be powerful when multiple chart components exhibit a common value domain~\cite{qu2017keeping}.
For instance, the green area chart in Figure \ref{fig:multichart}a---b shares a temporal domain with the stacked bar and pie charts (\ref{fig:multichart}a) and the orange area chart (\ref{fig:multichart}b).
The size and peripheral location of the green area chart and pie chart suggest that they are of secondary importance, providing context to the more visually prominent chart below it.
Accordingly, highlighting a point in time in the stacked bar chart can trigger corresponding highlights in the peripheral charts.
However, we allow presenters to override this linked highlighting (\ref{fig:multichart}b) by simultaneously pointing at different positions along the domains of two charts.

\subsection{Pinch and Drag to Transform and Compare}
\label{sec:ixd:manipulating}

    \begin{figure*}[h!]
        \includegraphics[width=\linewidth]{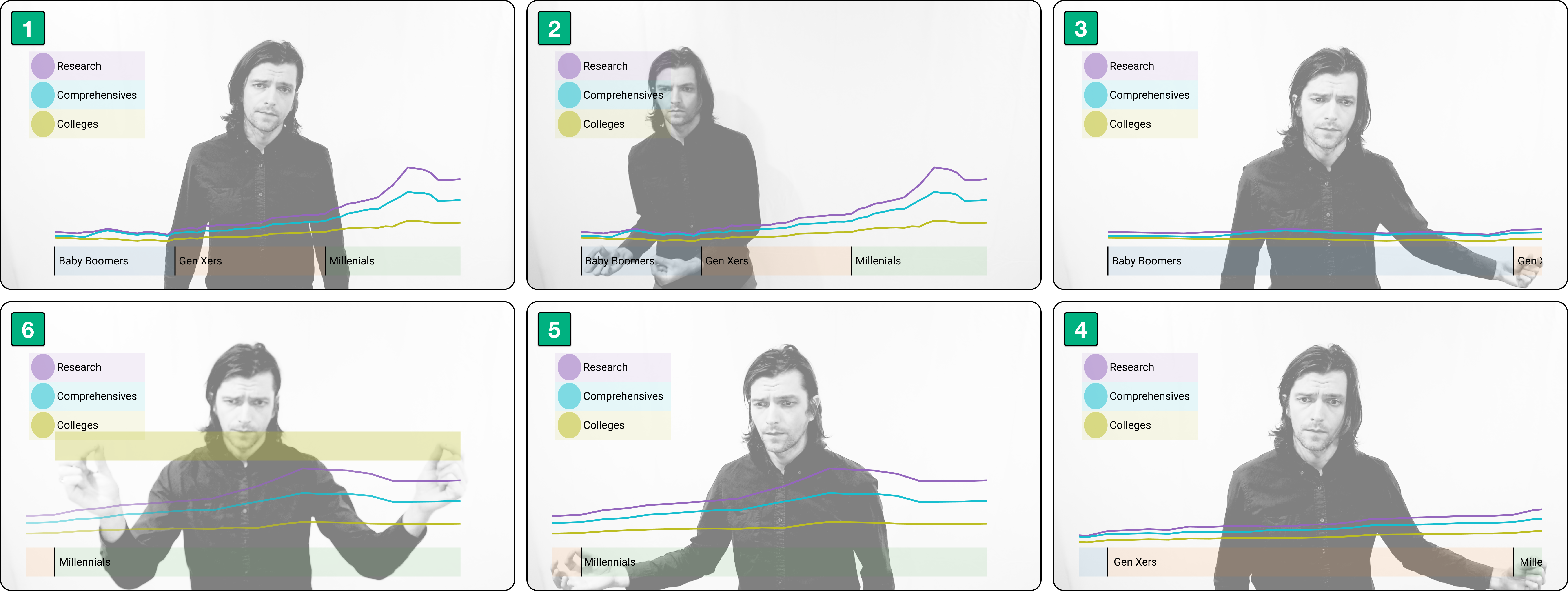}
        \vspace{-5mm}
        \caption{Manipulating the temporal scale of a chart where time appears on the X axis (clockwise starting from \#1): bimanual pinching in the bottom margin to zoom in on the timeline (2), pinching in a bottom corner margin to pan the timeline in either direction (3---5), and bimanual pinching in the top margin to zoom out on the timeline (6). The associated flourishes include \textit{expanding} when zooming in, \textit{flicking} or \textit{swiping} when panning, and \textit{compressing} when zooming out.}
        \label{fig:timeline}
        \vspace{-2.5mm}
    \end{figure*}

    \begin{figure*}[h!]
        \includegraphics[width=\linewidth]{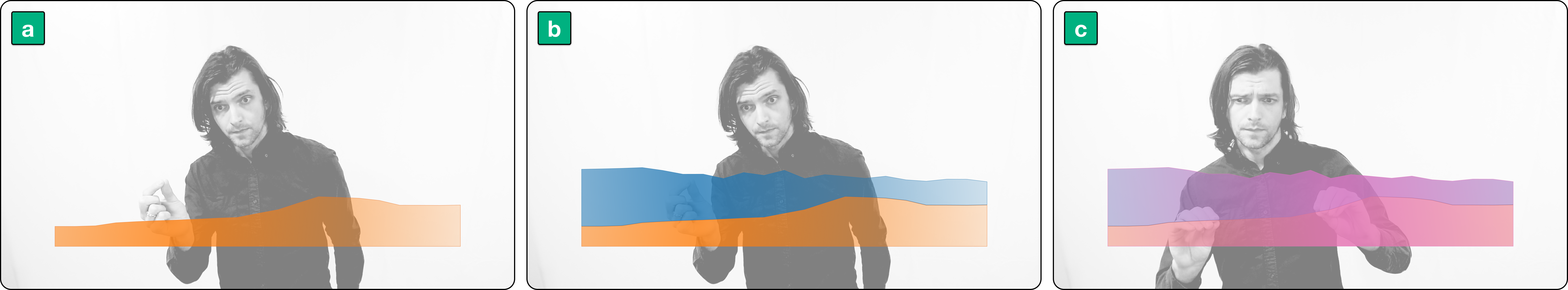}
        \vspace{-5mm}
        \caption{Pinching reveals a second band in the stacked area chart (a, b), where a possible flourish includes \textit{snapping} one's fingers; bimanual pinching ephemerally aggregates bands in the stacked area chart into a single band until the both pinches release, as shown (c), a presenter can \textit{crimp} the boundary between the two bands as a flourish.}
        \label{fig:snapcrimp}
        \vspace{-2.5mm}
    \end{figure*}

Figure \ref{fig:multichart}c---f illustrates the selecting and repositioning of individual chart elements. 
While the direct manipulation and dragging of individual chart elements may evoke the activities of chart construction~\cite{satyanarayan2019critical} and exploratory data analysis~\cite{saket2016visualization}, this category of interaction can also serve a communicative purpose.

\bstart{Illustrating a value transformation}
Consider how to explain a uniform transformation applied to a set of values, such as multiplying the GDP per capita of countries by their populations to derive total GDP values. 
While this should be explained in a presenter's oration, this explanation could also be reinforced by interaction.
We illustrate such a transformation in Figure \ref{fig:multichart}c---d, in which the presenter pinches to select the green area chart corresponding to student enrollment over time, generating a temporary copy (\ref{fig:multichart}c). 
In this example, the bar chart of individual student costs spans the same temporal domain, and in dragging and releasing the copied enrollment values over the bar chart, the copy is destroyed and the bar chart updates to reflect total cost values over time (\ref{fig:multichart}d).

\bstart{Comparisons on demand}
Another typical scenario is comparing values that are not spatially adjacent. 
While such comparisons could be planned in advance, we expect unplanned comparisons, such as those prompted by audience questions~\cite{brehmer2021jam}.
Although these unplanned comparisons could be performed by executing a filter command, this may result in a loss of context for the audience. 
We demonstrate an alternative way to compare elements in Figure \ref{fig:multichart}e---f, where each hand can pinch within the bounds of an element to create a cloned copy, one that is tethered to the position of the fingertips. 
As long as the pinches are maintained, these cloned elements can be freely repositioned (\ref{fig:multichart}e), so as to afford a side-by-side comparison of values.
Meanwhile, the original chart elements are de-emphasized apart from a stroke to indicate the source of the cloned elements.
Releasing the pinch destroys the cloned copies and restores the source chart to a normal opacity level (\ref{fig:multichart}f).


\subsection{Supporting Rhetorical Flourish}
\label{sec:ixd:flourish}


Some of the most prominent visual transformations that a presenter can apply to a chart include manipulations of scale and the addition or removal of chart content.
While animation can be a powerful cue to draw viewers' attention to these transformations~\cite{Heer2007}, a presenter's body language can provide complementary illustrative and affective cues.
However, the salience and deliberateness of a presenter's movement should be commensurate with the extent of visual transformation taking place, and with the transformation's relative importance with respect to the overall narrative.
Accordingly, several of the interactions supported by our presentation environment are quite deliberate relative to deictic highlighting.
Moreover, we give presenters the freedom to adjust the visual salience of these interactions; we describe this as the ability to add a \textit{rhetorical flourish} when performing the interactions.

\bstart{Express changes of temporal scale}
We demonstrate several opportunities for flourish when manipulating the temporal scale of a chart in Figure \ref{fig:timeline}, beginning with \ref{fig:timeline}.1 and proceeding clockwise.
This particular line chart of higher education tuition spans more than 50 years, with regions along the horizontal margin indicating when three generations attended college.
To zoom in on the Baby Boomer generation, for instance, the presenter pinches the corresponding span along the horizontal margin with both hands (\ref{fig:timeline}.2), which is a fairly deliberate and salient gesture.
Optionally, the presenter can pull their hands apart after pinching; this \textit{expansion} flourish does not affect the recognition of the interaction, but it visually reinforces that a zooming in is occurring.
The salience of this flourish is left to the discretion of the presenter: they could perform a small lateral expansion, or a more emphatic expansion heralding a more important reveal of information.
After zooming in, pinching the corner of the margin with one hand will trigger a panning of the timeline (\ref{fig:timeline}.3---\ref{fig:timeline}.5).
We allow for a degree of flourish here too, in that the presenter can \textit{flick} or \textit{swipe} their hand in the direction the pan, once again complementing the visual transformation taking place.
Lastly, bimanual pinching in the top margin or with a hand in both side margins will zoom out to restore the full timeline (\ref{fig:timeline}.6).
Mirroring the flourish for zooming in, pairing these pinches with a \textit{compression} gesture can draw additional attention to the change in temporal scale. 

\bstart{Revealing new chart elements}
Progressively unveiling highly salient chart elements offers another opportunity for flourish.
For instance, pinching within the bounds of an area chart can reveal an initially hidden second band (Figure \ref{fig:snapcrimp}a---b).
Pinching within the contours of the hidden band will draw more attention to the band as it appears, and pairing this pinch with an optional and audible finger \textit{snap} may draw even more attention.
Highly salient visual changes may also be ephemeral: for instance, bimanual pinching within the stacked area chart reveals a total value band superimposed over the individual bands (\ref{fig:snapcrimp}c), and this band remains in place as long as one hand is pinching within the bounds of the chart, leaving the other hand free to point out specific total values.
A possible flourish here is one in which the presenter assumes a bimanual \textit{crimping} pose along the boundary between the bands in the area chart, simultaneously satisfying the requirements of a bimanual pinch while visually reinforcing the aggregation of values.

\section{Evaluation}
\label{sec:eval}




We evaluated our approach to presenting data in two independent studies: one considering the perspective of the presenter and another considering the perspective of the remote audience.

\subsection{Procedure}
\label{sec:eval:procedure}

Irrespective of whether a participant was participating in the presenter study or audience study, we began each session by asking them about their current experiences with respect to attending and delivering remote presentations about data.
We then introduced the remote presentation scenario, which we described as an informal presentation among colleagues, one where the audience could interrupt the presenter and ask questions. 
The topic of the presentation pertained to higher education costs, for which we used data published by the Economic Opportunity Institute~\cite{tableaupubliceducation}.
At this point, the procedure diverged depending on the study, which we describe below.
We closed both session types with a reflection on the participant's experience, in which we asked them to extrapolate from our approach to consider other chart types and configurations, as well as other possible interactions that might support a presentation about data.
In both studies, sessions lasted between 45 and 60 minutes, which we recorded and transcribed.

\begin{figure}[h!]
        \vspace{-2.5mm}
        \includegraphics[width=\linewidth]{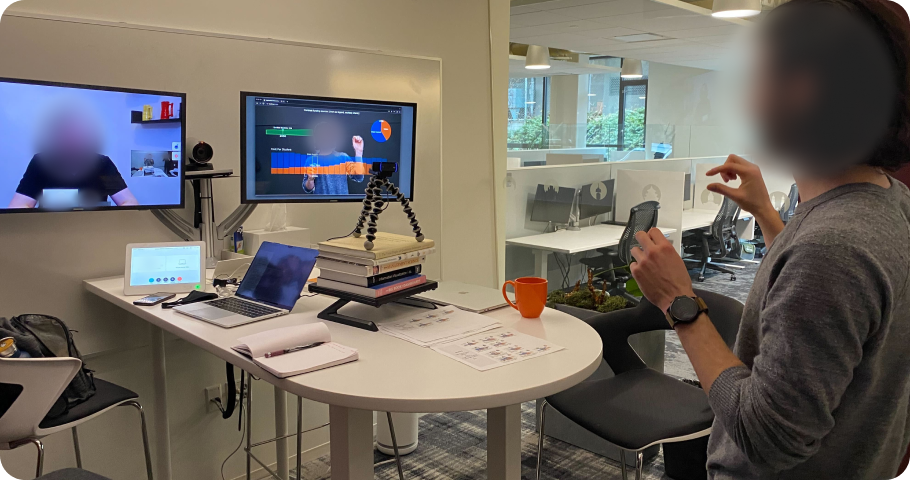}
        \vspace{-5mm}
        \caption{In the presenter evaluation, a webcam \revision{(mounted on a tripod placed at the center of the table)} captures a participant's hand motions while chart overlays on their video appear in the right monitor; a researcher assuming the role of the remote audience appears in the left monitor.}
        \label{fig:presenter}
        \vspace{-2.5mm}
    \end{figure}

\bstart{Presenter-oriented study}
The aim of the presenter-oriented study was to collect feedback on the utility, usability, and learnability of the presentation environment.
To familiarize presenter participants with the charts and their underlying data prior to their sessions, we directed them to a Tableau Public workbook~\cite{tableaupublic} containing a series of charts (featured in the supplemental video).
To ensure a consistent experience for participants, we observed them as they interacted with our presentation environment in a small meeting room in a corporate office setting (Figure \ref{fig:presenter}).
One researcher was present in the room with the participant, while another researcher assumed the role of a remote audience, with their video feed shown on one display. 
We displayed our presentation environment on another display and also shared it via a videoconferencing application. 
Finally, we pointed a Logitech 1080p webcam at the participant, positioning it to \revision{capture them from the waist upwards and to} allow for either a seated or standing presentation delivery, depending on their preference.
\revision{After toggling our environment's option to display fingertip control point markers,} we then progressed through eight scenes of a presentation, with each scene featuring different combinations of chart overlays and associated interactions.
For each scene, we invited participants to discover and practice the scene's associated interactions, and to think aloud as they interacted; we also provided them with a set of printed interaction reference sheets (adapted from Figures~\ref{fig:palmpointmargin} --- \ref{fig:snapcrimp}).
When participants struggled to perform a particular interaction, we provided verbal hints. 
Once familiar with the interactions in each scene, the remote researcher posed a question about the data, so as to serve as a presentation prompt, such as \textit{``can you describe how tuition costs changed for millennials during the 2010s?''}.

\bstart{Audience-oriented study}
The aim of the audience-oriented study was to better understand how presentations delivered using our environment might engage audiences.
In each session, one researcher took on the role of a presenter, while another researcher assumed the role of a meeting host or moderator. 
To better understand how a remote audience would experience a presentation about data delivered using our environment, we prepared two 5-minute presentations adapted from the same content used in the presenter-oriented study; we include abbreviated versions of these presentations in the supplemental video.
We additionally prepared alternate versions of these \revision{two} presentations \revision{that were more representative of current presentation practices~\cite{brehmer2021jam}; these retained} the same speaking points but we delivered them by screen-sharing the Tableau Public workbook~\cite{tableaupublic} used in the presenter-oriented study.
\revision{Unlike a slide presentation, a workbook allowed us to present content in unplanned ways, such as in response to audience questions.}


The presenter then delivered \revision{two of the four} presentations; to mitigate a potential novelty effect, half of the audience participants saw the alternate screen-share presentation \revision{followed by one} delivered using our presentation environment, while the other half experienced the reverse order.
Midway through each presentation, the moderator posed a clarifying question about the data in cases where the participant had not already interrupted or asked a question, so as to remind participants of the live presentation scenario.
After each presentation, we asked participants to summarize the content of the presentation and whether they had any content-related questions for the presenter.
\revision{As the interviews were open-ended, we found it useful with some participants to briefly demonstrate} what the alternate presentation delivered via screen-share would have looked like had it been \revision{delivered using} our presentation environment, \revision{so as to discuss specific points of contrast.}

\subsection{Participants}
\label{sec:eval:participants}

We recruited 29 professionals who attend, prepare, and deliver customer-facing presentations about data at a multinational software company, as well from its' subsidiary and customer organizations.
Our participants varied in terms of years of experience (from several months to over two decades) and job title (\eg solution engineer, product manager, account executive). 
While some participants were accustomed to presenting data to remote or mixed audiences prior to the COVID-19 pandemic, all of them have been presenting primarily to remote audiences since the pandemic's onset.
The majority of our participants reported attending multiple presentations a week involving some discussion about data, and all of our participants reported giving such presentations themselves, with cadences ranging from quarterly to daily.
For the presenter study (N$=$12), we limited our recruitment to those who were willing and able to meet with us in person at an office that required all employees to observe strict COVID-19 safety protocols.
For the audience study (N$=$17), participants joined remotely from locations spanning North America, South America, and Europe.
\subsection{Observations \& Participant Reflections}
\label{sec:eval:results}


We performed a thematic analysis of our observations and transcribed quotes from the session recordings. 
This process yielded several themes and tags, which we list in the supplemental material.
Throughout this section, participant attributions are study-specific, with P\# referring to a presenter study participant and A\# referring to an audience study participant.

\bstart{Reflecting on utility}
\revision{When we asked presenter} study participants to describe the potential value of adopting our approach for their own presentations.
P3 hypothesized that this would \textit{``probably lead to more discussion than a normal presentation,''} adding that audiences are \textit{``definitely not gonna forget,''} suggesting that seeing the presenter interact with data in this way will elicit more audience engagement, which will in turn result in a more memorable presentation experience.
As to what might drive this focused attention, P8 suggested that the format would force his audience \textit{``to look at the data closer, just by nature of me putting the effort into doing all this\ldots it would put an organic impetus on the audience\ldots it's like a circus, I'm juggling things --- you got to watch me, right?''} 
This focused audience attention could be particularly important when communicating causal relationships in data: P5 will start \textit{``with an overview and break it down into `here's what's causing that change'''}, and he saw our approach as being potentially \textit{``really helpful to explain and understand data at that level''}.

\bstart{A potential to engage}
From the audience perspective, participants were able to directly contrast our approach with the convention of presenting data via screen-sharing. 
In describing our approach, A5 said: \textit{``the storytelling is more engaging\ldots it connected more.''}
The overlay presentation also suspended A3's tendency to critique a presenter's delivery, likening it to instances where an audience unfamiliar with interactive business intelligence dashboards might similarly suspend their critical tendencies if they were accustomed to seeing presentations delivered primarily via slideware.
Also notable is A10's criticism of the screen-share presentation as lacking annotation and axis labelling, \revision{which we attribute to his tendency to associate business intelligence workbooks with data analysis, as opposed to synchronous presentation.} 
Following the overlay presentation, he did not offer the same criticism despite a similar absence of labelling: \textit{``I'm actually listening more, I'm paying attention more because I'm more engaged by your physical body, \ldots I like it better now that it's clean and not have all those numbers that I was looking for before\ldots in this version, I'm able to just listen to your voice and focus to the story\ldots my heart is more patient with you.''} 

For some audience members, their engagement was palpable: \textit{``I love this thing\ldots that gets my heart going a little bit''} [A12].
A4 summarized the value of this potential to increase audience engagement: \textit{``it's drawing attention because it's so engaging\ldots I actually want to understand what's going on behind the data, and that's where we want to be, because ultimately, data is for driving business decisions.''}
Given this guided attention and a realization that the presenter can interact directly with the content, A17 foresaw audiences asking more specific questions: \textit{``if I said: `hey, what is that peak over there?'\ldots you could just put your finger.''}

\bstart{A potential to distract}
Audience study participants were split as to how and when our approach to presenting data might distract audiences. 
On the one hand, A10 found to the experience to be neither confusing or distracting, and A4 described how \textit{``it pulls you in a good way where the technology doesn't become a distraction''}.
On the other hand, participants A16 found the presenter's body language distracting, while A8 worried that audiences would derail presentations by interrupting presenters with requests to perform more interactions.

Falling between these perspectives, A1 wavered on the potential of the interactions to distract audiences: \textit{``it sometimes feels natural and it sometimes feels stilted''}, while both A3 and A14 remarking on how their perspective shifted as the presentation proceeded, with A3 stating: \textit{``initially I wrote down `gimmick', because I felt like the effect of what you were doing was more distracting than the data I was getting, but that very quickly subsided\ldots I feel like I'm being guided through a story.''} 

To assess if audiences were distracted by our approach, we collected and contrasted their presentation content summaries, along with any content-related interruptions and questions.
Ultimately, we did not observe any pronounced differences in the quality of content summaries for a screen-share presentation and those for a presentation supported by interactive overlays, however the latter did elicit more substantive content-related interruptions and questions. 
Despite encouraging participants to assume informality and to interrupt the presenter with content-related questions, the majority of the interruptions made during the overlay-supported presentations pertained to \textit{how} the content was being presented.

\bstart{When not to use our approach}
There are teleconference scenarios involving visualization where our approach might not be applicable.
Both P4 and P11 described giving demonstrations of the functionality of data analysis software, where audiences already have an understanding of relevant analytical concepts.
P5 also described customer service meetings wherein he would present a dense dashboard to customers, meetings where there is already a shared context and an intent to resolve an issue, rather than an intent to reveal a narrative based in data.

\bstart{Natural and personal}
Most of our presenter study participants commented on how the experience of presenting data with interactive chart overlays made them feel relative to a conventional screen-sharing approach. 
P6 stated that \textit{``it feels more natural''} while also alluding to the agency or ownership that he felt with respect to the presentation content: \textit{``it's more obvious that I'm the one presenting this information\ldots} [the audience] \textit{can ask me the question, and not just ask a chart.''}
Similarly, P8 \textit{``felt more control over the data\ldots you feel like you're really telling the story with your whole body.''}
Continuing this theme, P5 described how the interface gave him more independence with respect to the language he would use when speaking to an audience: \textit{``I felt like I had more freedom,''} specifically referring to the use of tedious spatial cues such as \textit{``in the bottom left of this visualization.''}
Beyond personal agency, P1 described how the experience \textit{``felt more like teaching,''} highlighting the potential to apply our approach in situations where remote audiences require data literacy and graphicacy~\cite{balchin1966graphicacy} training.
Finally, P6 commented that our approach adds a human element to the seemingly precise domain of data: \textit{``it adds an emotional and personal element to something that's usually pretty right and wrong,} [where]~\textit{there's one logical answer.''}

In contrast to the comments about how our approach might make presentations about data feel more personal, P9 offered a different perspective based on his familiarity with livestreaming practices on YouTube and Twitch: overlay-supported presentations \textit{``would feel more professional''} and \textit{``performative''} than screen-shared slide presentations, and this would make them more engaging.
On the other hand, P10 felt \textit{``like you're playing more with the data,''} suggesting a more casual or informal presentation experience.

\bstart{Flourishes and interaction variations}
The majority of audience study participants made references to specific interactions performed by the presenter in presentations supported with interactive overlays.
In particular, pinching to clone and drag chart components for value transformation or comparison (Figure \ref{fig:multichart}c---\ref{fig:multichart}f) was mentioned by six participants, with A13 remarking how difficult it would be to perform these interactions using conventional presentation approaches.

We also heard comments pertaining to how some of our interactions allowed for flourish or embellishment without unintended effects on the content.
A9, A12, and A15 each singled out the \textit{snap} to reveal a band in the stacked area chart (Figure \ref{fig:snapcrimp}a---\ref{fig:snapcrimp}b), despite acknowledging that this reveal was triggered by a simple pinch. 
A12 additionally praised \textit{swipe} flourish when panning the timeline of the line chart (Figure \ref{fig:timeline}.3---\ref{fig:timeline}.5) and a \textit{bloom} gesture that the presenter performed while the stacked area chart transitioned into a stacked bar chart (see supplemental video), describing these as \textit{``jazz hand maneuvers\ldots the theatrical performing part of my brain did really enjoy that.''}
From the presenter perspective, P8 specifically called out the allowance for flourish when panning the timeline of a chart: \textit{``it's like on the iPhone\ldots give it a little wrist and it goes\ldots you feel like you're more connected with it.''} 

One unanticipated reflection on the interactions came from A9, who routinely attends and leads presentations with his customers in the public sector, though he specifically spoke about his customers in law enforcement and the armed forces. 
He mentioned how in these domains, mannerisms such as single-digit pointing are discouraged in interpersonal communication, as they have been found to elicit a heightened stress response.
As our deictic interactions can accommodate open-handed pointing, this alleviated his concerns regarding the viability of our approach in these domains.

\bstart{Reflecting on usability}
Many presenter study participants initially assumed awkward or uncomfortable poses, such as stiffly pointing straight at the camera or raising their elbow to be level with their hand.
However, within minutes, they assumed more comfortable poses.
Retrospectively, several participants appreciated being able to interact with content using their hands, describing it as \textit{`comfortable'} [P7] and \textit{``more effective than speaking to something and using your mouse to point out stuff in a presentation''} [P3]. 

On the other hand, P11 seemed less comfortable: \textit{``I'm using my whole body, which is uncomfortable\ldots maybe just because it's different, not because it's bad.''} 
P12 also expressed concern that \textit{``some of the natural hand gestures that we're using for speaking could be confusing for this technology,''} which certainly resonated with us given our intention to avoid any such collisions. 

Specific interactions also elicited some concerns, such as P6's need for precision pointing: \textit{``the advantage of having a cursor or track pad is that it's a little easier to know exactly where you're going and take your hand off the mouse\ldots if you want something to stop, you can't really cut your hand off and make the visual pause,''} highlighting a need to disable and enable hand-tracking on demand.
Aside from pointing, the pinch-based interactions were also an occasional source of frustration; P3 voiced a concern about testing her audience's patience should she require multiple attempts at executing a pinch, and while P7 stated that \textit{``not many things were challenging,''} he found it difficult to know \textit{``how to keep my hand away from the camera so that it feeds it properly.''}
Lastly, pinching proved especially difficult for P12, whose dark nail polish appeared to thwart hand pose recognition.

\bstart{Speculating about learnability}
All but one of the presenter study participants commented on the process of learning how to effectively interact with chart overlays.
An overall sense of optimism was notable, such as when P4 stated \textit{``it's a learning process; that doesn't bother me\ldots I actually really enjoy figuring this thing out and watching it respond.''}
However, as P9 points out, our approach could be \textit{``a lot to figure out on the fly,''}, with P12 suggesting that presenters might benefit from a \textit{``significant amount of training before''} giving their first presentation to an audience. 
Beyond individual discovery of the interactions, P3 adds: \textit{``I think watching someone else do it first is probably very critical for training.''}
Despite a perceived learning curve, we were encouraged by P5 and P8 describing our interactions as \textit{`intuitive'}, and as with any new approach, P6 stated: \textit{``it probably just takes some getting used to.''}

Audience participants also speculated on learnability, expressing a similar blend of optimism and concern with respect to an expected learning curve, particularly if we contrast A5 imagining a time when \textit{``we all learn these skills --- I presume that's not a challenge''} with A16 suggesting that the presenter had extensively practiced the presentation beforehand.




\subsection{Eliciting Ideas to Extend our Approach}
\label{sec:eval:elicitation}

When asked about what is uniquely challenging or frustrating about presenting data to remote audiences, two themes stood out: the ability to engage an audience whose attention is divided and difficulties with presenting either complex visuals or complex data. 
We now consider the ideas we elicited from participants that speak to these themes as a basis for extending our approach.

\bstart{Engaging an audience and managing attention}
While many audience study participants reported feeling engaged and focused on a dynamic presenter, the interaction with overlays was ephemeral, making it difficult for audiences to attend to highlighted values and their related insights.
Meanwhile, presenter study participants P2, P7, and P9 independently expressed a desire to pin value annotations in place; \textit{``I instinctively wanted to push it hard and stick there''} [P2], which suggests a potential for gesture recognition at various depths from the camera, and recalls a similar interaction demonstrated by Gong~\etal~in their HoloBoard presentation system~\cite{gong2021holoboard}.
While a purely functional \textit{pin} gesture like this may seem natural for presenters, we are unsure of how it might be interpreted by audiences.
This potential mismatch calls for a deeper exploration of \textit{pinning} and \textit{unpinning} interactions for data marks and annotations from the perspectives of both presenter and audience.

We also foresee a need for unplanned annotations to be added during a presentation.
Such functionality could encourage audiences to volunteer their own insights on the data; a presenter could capture and externalize these insights, perhaps through speech-to-text annotation [P6], binding these annotations to data marks based on hand proximity or via linked highlighting (Section \ref{sec:ixd:multiple}).
Alternatively, A6 suggested that a presenter could pass interactive privileges to an audience member: \textit{``when you picked up one of those [bars], you could have said `do you want to try and move it?'\ldots And then I could have moved it to show you what I'm talking about.''}

\bstart{Presenting complex data}
While bar, line, area, and pie charts may satisfy the needs of many presentations of data~\cite{brehmer2021jam}, 16 of our participants urged us to consider a wider palette of data and chart types, which could in turn expand our repertoire of interactions.
Specifically, multiple participants mentioned scatterplots, both symbol and choropleth maps, richly-formatted text tables, treemaps, and representations of distributions such as boxplots.
However, we appreciate P12's sense of caution: \textit{``it's going to be difficult\ldots especially when you get into very detailed charts,''} suggesting the need for a different approach to interaction for charts with a greater number or density of elements. 


Beyond a wider palette of supported charts, A4 and A16 urged us to consider non-linear narratives and the ability to break out of a planned sequence or arrangement of charts in response to audience engagement.
Whether such scene management and content transformation functionality are activated by introducing a new set of non-distracting gestures or through off-screen presenter controls is a question worthy of further research and design.

\section{Discussion}
\label{sec:discussion}
We now reflect on the potential of our approach, proposing design implications for presentation tools that reinforce and extend previous research.
In Section \ref{sec:proc}, we stated our goal of improving the data presentation experience for both presenter and audience, with two design imperatives: (1) to support direct interaction with visualization content rather than interaction with the containing scene or slide; and (2) to realize a presentation environment that would require no special equipment and leverage an existing fluency in communicative body language encompassing illustrative and affective hand gestures.
In developing our environment, the interplay between these imperatives made us aware of how introducing novel functional interactions can be an imposition on presenters as well as a potential source of distraction for audiences.
As a result, we focused on identifying interactions that were simultaneously functional and deictic, interactions that could be embellished with rhetorical flourishes.


    \begin{figure}[h!]
        \includegraphics[width=\linewidth]{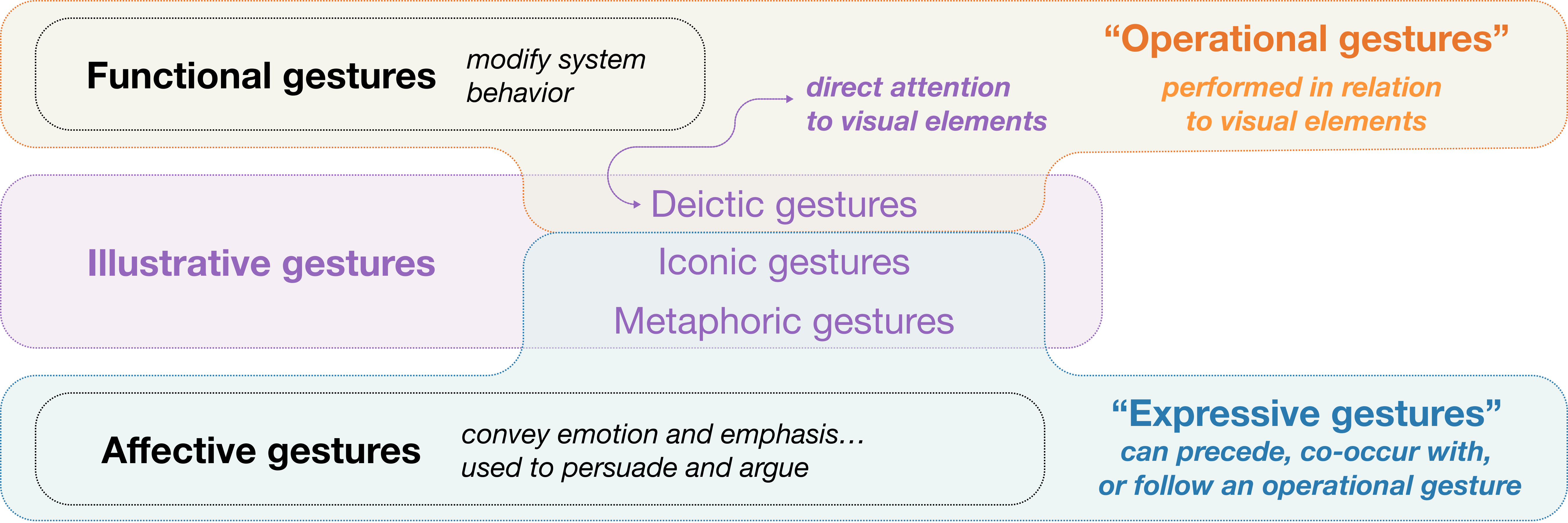}
        \vspace{-5mm}
        \caption{Our pragmatic reclassification of functional, illustrative, and affective gestures for presenting data, which distinguishes between operational gestures performed in relation to specific visual elements and expressive gestures that can be performed independent of any visual aid.}
        \label{fig:gesture}
        \vspace{-2.5mm}
    \end{figure}

By reflecting on these design imperatives, we now have a better understanding of what hand gestures are useful when talking about data with interactive chart overlays.
\revision{In particular, we can} revisit the categorization of hand gestures described in Section \ref{sec:rw} from a pragmatic perspective (Figure \ref{fig:gesture}).
Deictic and functional gestures can be considered to be \textit{operational} in that they mediate the experience of attending to dynamic visual artifacts such as interactive chart overlays.
In contrast, iconic, metaphorical, and affective gestures facilitate \textit{expressive} human-to-human communication; they are an essential component of rhetoric, and they can be performed independent of any visual aid.
With this new understanding, we propose three design implications for presentation tools that support touchless interaction with data: 
\bstart{Recognize operational gestures}
We contend that the most fruitful potential path lies in recognizing operational gestures performed in the context of dynamic charts, and this recognition should be independent of any affective interpretation. 
As a comparison, consider the recent presentation authoring system described by Saquib~\etal\cite{saquib_interactive_2019}, one capable of recognizing individual presenters' idiosyncratic mix of illustrative and affective gesture.
When speaking about familiar data abstractions and supported by a limited palette of associated charts, a presenter should be able to reliably perform a minimum set of operational gestures, each with an allowable tolerance in terms of how it is recognized.

Operational gestures should be familiar and leverage cultural norms, such as deictic \revision{pointing} gestures in interpersonal conversation or functional gestures learned from using ubiquitous touchscreen devices~\cite{cirelli_survey_2014}, \revision{such as swiping to pan along a continuous dimension}. 
Familiarity with operational gestures could also explain participants' optimism with respect to learnability; although many participants expected that learning how to use a presentation environment integrating our approach would take time and practice, \revision{we also heard our operational gestures described} as \textit{`intuitive'}.   


\bstart{Assign gestures \revision{according to frequency and salience}}  
As intuitive as some operational gestures may be, presenters likely need time to learn how to adapt these familiar gestures to a new medium, so system designers must be judicious when mapping operational gestures to interface functions.
Consider the functional pinch gesture: easy to perform on a touchscreen, but as we saw in the presenter study, it can be harder to perform with webcam-based hand tracking.
System response should also be commensurate with the deliberateness and expected frequency of the operational gesture.
For instance, single-handed pointing at elements in a chart can happen quite often during a presentation; in response, our environment reveals unobtrusive and ephemeral value annotations.
In contrast, pinching to select a chart element is a more deliberate and infrequent operational gesture, and our system responses are accordingly more overt (\eg zooming, panning, element cloning).


\bstart{Do no harm: avoid conflict with expressive gestures}  
A challenge and opportunity for future presentation systems is to encourage the richness of expressive body language without compromising system behaviour. 
In other words, the expressive gestures that a system ignores are just as important as the operational gestures that it acts upon.
While expressive gesture can be performed independently of an operational gesture, we demonstrated how an expressive gesture can immediately precede, follow, or take place concurrently with an operational gesture.
For instance, in Section \ref{sec:ixd:flourish}, we described \textit{expand}, \textit{swipe}, and \textit{compress} gestures as flourishes that follow an operational pinch gesture; the pinch triggered the system response of zooming or panning along the temporal domain of a line chart, while the flourishes helped to express the concept of a continuum of time, complementing the visual transition without triggering their own system response.
Consider that these iconic gestures could have supported verbal statements about different time intervals without the line chart even being shown.
Similarly, we described the affective \textit{snap} gesture as a rhetorical flourish following a pinch that triggered the reveal of a new chart element; here the snap both visually and audibly emphasized the importance of the newly revealed content.
Distinguishing expressive and operational gestures will undoubtedly present challenges, though a system could make inferences by collecting additional movement data, such as the acceleration or beat frequency of a fingertip or hand, as well as additional position data, such as the location of the hand relative to the body.

\bstart{Limitations and future work}
In our studies, we limited recruitment to those associated with a multinational software company, and given the office setting of the presenter study, COVID-19 safety protocols prohibited us from inviting non-employee participants. 
We hesitated to conduct the presenter study remotely, as we could not guarantee a consistent environment free from distractions in presenters' homes or remote co-working spaces.
\revision{On the other hand, our meeting room setting (Figure \ref{fig:presenter}) may have been potentially unrepresentative with respect to the relative positioning of the presenter, the webcam, and the displays.}
It was also challenging for participants to remain in character during the presentation scenarios. 
In the presenter study, we reminded participants of the scenario with prompts related to the presentation content, while in the audience study, we noted several presentation interruptions focusing not on the content but on the mode of delivery and the technology in use.
Whether audiences will engage the presenter with meaningful content-focused discussion and retain key messages should be examined in a longitudinal deployment study of a presentation system integrating our approach.

We identified several opportunities to extend our approach, though two of the recurring categories of ideas elicited by our study participants stand out.

First, we can identify additional ways to capture attention and increase engagement.
\revision{This includes allowing} the audience to interact with and annotate the content during a presentation~\cite{chung2021,willett2012strategies}, or provide a shared awareness of what audience members are looking or pointing at~\cite{gronbaek2021mirrorblender,he2021gazechat}.
\revision{Alternatively, we could explore more flexible and expressive chart annotation options for presenters by recognizing handheld peripherals such as pens or pointing devices, drawing inspiration from Perlin~\etal's Chalktalk~\cite{Perlin2017}}.

Second, \revision{we recognize that a palette of basic palette of charts will satisfy many presentation scenarios~\cite{brehmer2021jam}, but we must nevertheless} consider how to interact with more detailed or unfamiliar communication-oriented charts~\cite{kosara2016presentation}.
This could involve incrementally constructing a detailed chart through a series of operational gestures, or transforming~\cite{brosz2013transmogrification,ruchikachorn2015learning} a more familiar chart into a new configuration, such as through bending, stretching, rolling, or tearing chart components.

\section{Conclusion}
\label{sec:conclusion}

The capacity to perform communicative and collaborative knowledge work at distance has increased in recent years, particularly since the onset of the COVID-19 pandemic.
At the same time, organizations both large and small have acknowledged the importance of making decisions and adopting policies that are grounded in data.
Despite a need to discuss data with remote stakeholders, colleagues, and customers, existing tools for presenting data to these audiences fall short: presentations fail to engage audiences and the multimodal expressivity of presenters goes unseen. 

\revision{In this paper, we described an approach that aims to help} restore the multimodal richness and nuance of co-located presentations about data.
\revision{We demonstrated a presentation environment} that composites presenter webcam video with interactive visualization overlays that respond to a presenter's hand movements.
We identified a set of interactions that can support presentations about data, using them with different types of charts to draw an audience's attention to differences in quantity, proportion, and interval. 
We \revision{evaluated our presentation environment in two studies}, examining the perspectives of 12 presenters and 17 audience members \revision{with respect to utility, usability, and the capacity to engage or distract audiences.}
\revision{Ultimately, we remain optimistic about the potential of multimodal presentation tools for presenting data to remote audience.1
However, as tool builders expand the vocabulary of operational gestures applicable across chart types, we urge them to avoid potential conflicts with expressive gestures.}





\end{document}